\documentclass[letterpaper, 10pt, journal, twocolumn]{IEEEtran} 
\usepackage{graphicx}
\usepackage{winsnotation}
\usepackage{amsmath}
\usepackage{amssymb}
\usepackage{cite}
\usepackage{caption}
\usepackage{url}
\usepackage{eqparbox}
\usepackage{subcaption}

\usepackage{graphicx}
\usepackage{amssymb}
\usepackage{epstopdf}
\usepackage{multirow}
\usepackage{color}
\usepackage{colortbl}
\usepackage{xcolor}
\usepackage{mathrsfs}

\usepackage[top=.90in, bottom=1.1in, left=0.70in, right=0.70in]{geometry}
\newcommand{\keywords}[1]{\textbf{\textit{Index terms---}} #1}

\title{\LARGE \bf Self-supervised Contrastive Learning for 6G UM-MIMO THz Communications: Improving Robustness Under Imperfect CSI
\thanks{The fundamental research described in this paper was supported, in part,
by the National Science Foundation under Grant CNS-2148251 and
by federal agency and industry partners in the RINGS program.}}

\author{
\IEEEauthorblockN{Rafid Umayer~Murshed\IEEEauthorrefmark{1}, Md Saheed~Ullah\IEEEauthorrefmark{2}, Mohammad Saquib\IEEEauthorrefmark{1}, and~Moe Z.~Win\IEEEauthorrefmark{3}}
\\
\IEEEauthorblockA{Email: rafidumayer.murshed@utdallas.edu, saheed@udel.edu, saquib@utdallas.edu, moewin@mit.edu}
\\
\IEEEauthorblockA{\IEEEauthorrefmark{1}Department of Electrical and Computer Engineering, The University of Texas at Dallas, Texas, USA}
\\
\IEEEauthorblockA{\IEEEauthorrefmark{2}Department of Electrical and Computer Engineering, University of Delaware, Delaware, USA}
\\
\IEEEauthorblockA{\IEEEauthorrefmark{3}Laboratory for Information and Decision Systems, Massachusetts Institute of Technology, Massachusetts, USA}
\vspace{-3.4ex}
}

\begin{document}

\maketitle
\thispagestyle{empty}
\pagestyle{empty}

\begin{abstract}
This paper investigates the potential of contrastive learning in 6G ultra-massive multiple-input multiple-output (UM-MIMO) communication systems, specifically focusing on hybrid beamforming under imperfect channel state information (CSI) conditions at THz. UM-MIMO systems are promising for future 6G wireless communication networks due to their high spectral efficiency and capacity. The accuracy of CSI significantly influences the performance of UM-MIMO systems. However, acquiring perfect CSI is challenging due to various practical constraints such as channel estimation errors, feedback delays, and hardware imperfections. To address this issue, we propose a novel self-supervised contrastive learning-based approach for hybrid beamforming, which is robust against imperfect CSI. We demonstrate the power of contrastive learning to tackle the challenges posed by imperfect CSI and show that our proposed method results in improved system performance in terms of achievable rate compared to traditional methods.
\end{abstract}


\section{INTRODUCTION}
\IEEEPARstart{U}{ltra-massive} multiple-input multiple-output (UM-MIMO) systems are a promising solution to meet the increasing demand for high data rates and reliable communication in next-generation 6G wireless networks. By utilizing a large number of antennas at both the transmitter and receiver, UM-MIMO systems can achieve significant improvements in spectral efficiency (SE) and capacity\cite{akyildiz2016realizing}. However, these benefits come with the challenge of obtaining accurate channel state information (CSI) for designing effective beamforming strategies. 

Hybrid beamforming, which combines digital and analog beamforming techniques, is a widely used approach in UM-MIMO systems to reduce hardware complexity and power consumption\cite{ullahspectral,molisch2017hybrid}. By capitalizing on the strengths of digital and analog elements, hybrid beamforming adeptly manages the trade-off between performance and resource efficiency. This approach optimizes the overall SE of communication systems and addresses the challenges arising from hardware limitations in emerging technologies. However, the performance of hybrid beamforming is highly dependent on accurate CSI\cite{liu2016impact}. In practice, acquiring perfect CSI is challenging due to various factors such as channel estimation errors\cite{gao2023spatially}, feedback delays\cite{ghourab2023secure}, and quantization errors\cite{chang2021deep}. Therefore, there is a need for robust beamforming techniques that can perform well even with imperfect CSI. In the context of hybrid beamforming's reliance on accurate CSI, recent studies underscore the significance of considering channel estimation inaccuracies. For instance, the work on diversity combining reveals how systems can retain diversity order despite imperfect channel estimates in correlated fading environments\cite{Ref1DiversityCombining}. Similarly, investigations into antenna subset diversity illustrate the performance of systems employing subset selection under non-ideal channel estimation, maintaining the diversity order while highlighting the impact of estimation errors on system efficiency\cite{Ref2AntennaSubset}. These observations highlight the necessity of implementing robust beamforming solutions that factor in imperfect CSI to enhance UM-MIMO system performance.

Harnessing the power of the NYUSIM channel model, an advanced simulation tool from New York University, we can accurately emulate the complexities of real-world wireless environments. NYUSIM allows for simulations with RF bandwidths up to 1 GHz and carrier frequencies reaching 150 GHz, exceeding the scope of 5G technologies \cite{poddar2023tutorial}. It offers a versatile platform for testing diverse scenarios like urban microcell (UMi) and macrocell (UMa), as well as indoor hotspots (InH), through adjustable parameters such as spatial consistency and human blockage \cite{ju2018millimeter}. These features enable a detailed replication of the multifaceted nature of THz wireless communication environments, making NYUSIM an invaluable asset for researchers and engineers examining the intricacies of wireless systems \cite{murshed2024fast}.

In light of the above, our paper introduces a pioneering self-supervised contrastive learning-based approach for hybrid beamforming in UM-MIMO THz systems, particularly under imperfect CSI. Contrastive learning has demonstrated remarkable success in diverse domains such as computer vision, natural language processing, and even seismology\cite{Murshed_TGRS}, to discern and exploit the subtle nuances in data. We meticulously harness self-supervised contrastive learning to address imperfect CSI, leveraging its ability to enhance robustness by identifying subtle differences in beamforming configurations amidst noise. By exploiting UM-MIMO's high spatial resolution, we effectively counteract CSI inaccuracies, which is particularly crucial due to these systems' line-of-sight (LOS) path predominance. Furthermore, the UM-MIMO architecture's fewer multipath components and the precise beams formed by extensive antenna arrays necessitate unparalleled accuracy in beamforming strategies. This precision becomes challenging for existing algorithms \cite{yu2016alternating} under noisy CSI conditions. Our selection of contrastive learning is influenced by its proven capability to discern subtle data nuances under such constraints.

\emph{Notation:} Scalars are denoted by lowercase letters, while vectors and matrices are represented by bold lowercase and uppercase letters. The absolute value, $L_2$ norm, and Frobenius norm are symbolized as $|\cdot|$, $||\cdot||$, and $||\cdot||_F$, correspondingly, with the Hermitian transpose indicated by $\dag$. The identity matrix of size $m \times m$ is expressed as \(\M{I}_{m}\). Additionally, the expected value, complex number set, and natural number set are designated by $\E{\cdot}$, $\mathbb{C}$, and $\mathbb{N}$, in that order, ensuring mathematical rigor and coherence throughout our analysis.

\section{System Model and Problem Formulation}
This section presents the system model for UM-MIMO communications and formulates the hybrid beamforming problem under imperfect CSI conditions. We discuss the challenges associated with conventional hybrid beamforming techniques and motivate the use of contrastive learning to address these challenges.

\subsection{System Model}
Our single-user UM-MIMO network features a base station (BS) and user equipment (UE), each with a uniform linear array (ULA) of \(N_{t}\) and \(N_{r}\) antennas, respectively, and \(N^t_{\mathrm{RF}}\) and \(N^r_{\mathrm{RF}}\) RF chains for transmission and reception. The network supports \(N_{s}\) data streams, constrained by \(N_{s} \leqslant \min(N^t_{\mathrm{RF}}, N^r_{\mathrm{RF}})\). The limited RF chains necessitate hybrid digital-analog beamforming, with the transmitted signal \(\RV{x} = \RM{F}_{\mathrm{RF}} \RM{F}_{\mathrm{BB}} \RV{s}\), where \(\RM{F}_{\mathrm{BB}}\) $\in \mathbb{C}^{N^t_{\mathrm{RF}} \times N_{s}}$ and \(\RM{F}_{\mathrm{RF}}\) $\in \mathbb{C}^{N_t \times N^t_{\mathrm{RF}}}$ are the digital and analog precoders, and \(\RV{s}\) is the transmitted symbol vector with \(\mathbb{E}[\RV{s}\RV{s}^{\mathsf{\dag}}] = \frac{1}{N_s}\M{I}_{N_s}\). The received signal can be modeled as
\begin{equation}
\RV{r} = \sqrt{\rho} \RM{W}_{\mathrm{BB}}^{\mathsf{\dag}} \RM{W}_{\mathrm{RF}}^{\mathsf{\dag}} \RM{H} \RV{x} + \RM{W}_{\mathrm{BB}}^{\mathsf{\dag}} \RM{W}_{\mathrm{RF}}^{\mathsf{\dag}} \RV{n}
\end{equation}
with \(\RM{W}_{\mathrm{RF}}\) $\in \mathbb{C}^{N_r \times N^r_{\mathrm{RF}}}$ and \(\RM{W}_{\mathrm{BB}}\) $\in \mathbb{C}^{N^r_{\mathrm{RF}} \times N_{s}}$ as the receiver's beamformers, \(\rho\) as transmit power, \(\RM{H}\) $\in \mathbb{C}^{N_t\times N_r}$ as the channel matrix, and \(\RV{n}\sim \mathcal{C}\mathcal{N}(0, \sigma^2\M{I}_{N_r})\) denotes the additive white Gaussian noise (AWGN) vector.

\subsection{Problem Formulation}
The beamforming challenge in UM-MIMO systems is twofold: precoder and decoder design. In this work, we demonstrate the application and potential of self-supervised contrastive learning to enhance robustness in UM-MIMO communications. Specifically, we concentrate on designing the digital part of the beamforming process under the assumption of a perfectly designed analog beamformer. This approach leverages the nuanced potential of ultra-massive MIMO systems, even in dominant LOS conditions, where the vast antenna array magnifies both capacity and the necessity for precise digital beamforming strategies to maximize SE.

Given this context, the precoder design can be formulated as
\vspace{-2ex}
\begin{subequations}
\begin{IEEEeqnarray}{RCL}
    & \underset{\RM{F}_{\mathrm{RF}}, \RM{F}_{\mathrm{BB}}}{\text{minimize}}
    & \qquad \left\| \RM{F}_{\mathrm{opt}} - \RM{F}_{\mathrm{RF}} \RM{F}_{\mathrm{BB}} \right\|_F \label{eq:objective} \\*
    & \text{subject to}
    & \qquad \left|(\RM{F}_{\mathrm{RF}})_{ij}\right| = 1\hspace{1.0cm} \forall i, j \label{eq:constraint1} \\*
    && \qquad \left\| \RM{F}_{\mathrm{RF}} \RM{F}_{\mathrm{BB}} \right\|_F^2 = N_s. \label{eq:constraint2}
\IEEEeqnarraynumspace
\label{eq:beamforming_optimization}
\end{IEEEeqnarray}
\end{subequations}
Maximizing SE is analogous to the above formulation \cite{yu2016alternating}. The optimal hybrid precoders closely resemble fully digital precoders in an unconstrained setting. Utilizing a MIMO channel's decomposition into parallel independent channels\cite{goldsmith_2005}, the data rate increases proportionally to the number of independent streams, \(N_{s}\). The optimal fully digital precoder matrix \(\RM{F}_{\mathrm{opt}} \in \mathbb{R}^{N_{t} \times N_{s}}\) is derived from the Singular Value Decomposition (SVD) of the channel matrix \(\RM{H}\) as \(\RM{H} = \RM{U} \Sigma \RM{V}^{\mathsf{\dag}}\)\cite{SVD_explanation}. In this context, the challenge lies in the absence of direct access to the noise-free optimal fully digital precoder matrix, \(\RM{F}_{\mathrm{opt}}\). Our formulation must contend with \(\Tilde{\RM{F}}_{\mathrm{opt}}\), a noise-corrupted approximation of \(\RM{F}_{\mathrm{opt}}\), resulting from the inevitable imperfections in CSI. This reality necessitates a robust design strategy that can effectively approximate \(\RM{F}_{\mathrm{opt}}\) and sustain performance despite CSI inaccuracies, underlining the critical need for resilient optimization in the digital beamforming domain. Upon establishing the digital beamforming matrix, the corresponding analog beamformer can be determined through a supervised approach, as elucidated in \cite{Raf_HBF}.

The SE is calculated as
\begin{equation}
R = \log_{2} \left| \M{I}_{N_s} + \frac{\rho}{\sigma^2N_s} \RM{W}_t^{\mathsf{\dag}} \RM{H} \RM{V}_t \RM{V}_t^{\mathsf{\dag}} \RM{H}^{\mathsf{\dag}} \RM{W}_t\right|
\end{equation}
where \(\RM{W}_t = \RM{W}_{\mathrm{RF}} \RM{W}_{\mathrm{BB}}\) and \(\RM{V}_t = \RM{F}_{\mathrm{RF}} \RM{F}_{\mathrm{BB}}\).

\begin{figure*}[!ht]
\centering
    \includegraphics[width=1.9\columnwidth,height=77mm,trim={3cm 3.5cm 5cm 3.4cm},clip]{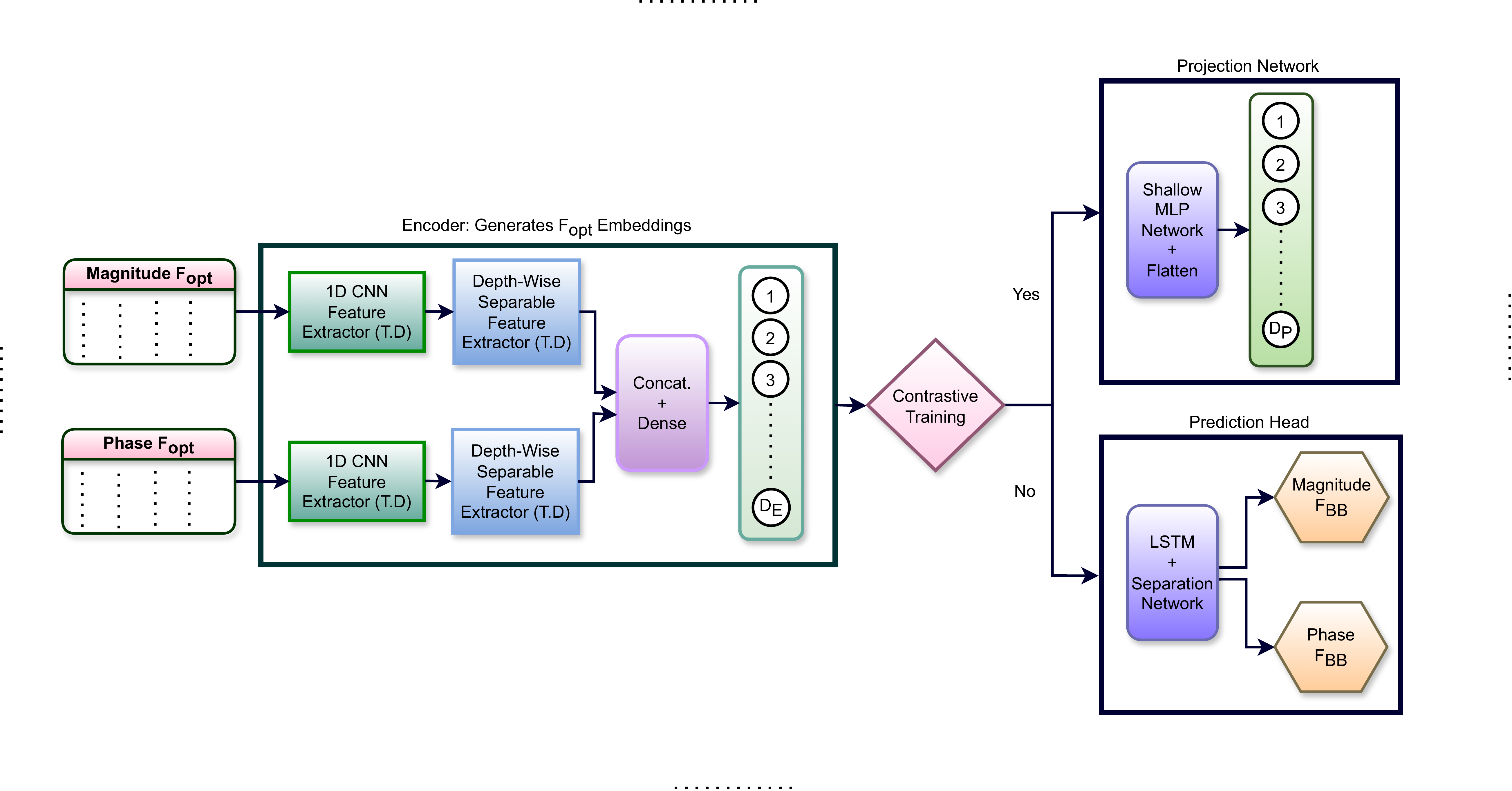}
    \caption{Schematic diagram of the proposed self-supervised contrastive DNN illustrating its architecture and workflow.}
    \vspace{-2mm}
    \label{fig:cont_block}
\end{figure*}

\section{Contrastive Learning-Based Approach}
This section describes our proposed contrastive learning-based approach for hybrid beamforming in UM-MIMO systems under imperfect CSI. First, the notion of contrastive learning is presented, and its application in the realm of hybrid beamforming is illustrated. We then present the architecture of our proposed deep learning model and describe the training procedure and optimization techniques used to learn the optimal beamforming weights.

\subsection{Contrastive Learning - Background}

Contrastive learning is a self-supervised learning paradigm that learns to encode data by contrasting positive pairs against negative pairs. Fundamental to this is the contrastive loss function, often formulated as a function of temperature-scaled dot products between representations. Specifically, for a pair of representations \( \RV{u} \) and \( \RV{v} \), the loss can be described as:
\begin{equation}
L_{\mathrm{cl}}(\RV{u}, \RV{v}) = -\log \frac{\exp(\RV{u^{\dag}} \RV{v} / \tau)}{\sum_{\RV{w} \in \Set{W}} \exp(\RV{u^{\dag}} \RV{w} / \tau)}. 
\end{equation}
Here, \( \Set{W} \) contains one positive sample and several negative samples, and \( \tau \) represents the temperature parameter, adjusting the sensitivity of the loss function to the similarity between samples. We set $\tau = 0.1$, a typical value that balances training stability with the capacity to generate robust embeddings \cite{khosla2020supervised}. This loss encourages the network to learn an embedding space where positive samples are pulled closer and negative samples are pushed apart. In our context of UM-MIMO systems and contrastive learning, positive pairs represent samples of beamforming matrices that are augmented versions of the same original matrix. This reflects similar channel conditions or beamforming solutions. These pairs teach the model to recognize and align similar patterns or conditions. Negative pairs, conversely, consist of samples from different channel conditions or beamforming solutions. These ensure the model distinguishes between fundamentally different scenarios, enhancing its ability to generalize across varied channel states and noise conditions.

\begin{figure}
     \centering
     \begin{subfigure}[b]{0.5\textwidth}
      \centering
        \includegraphics[trim={1.5cm 0.1cm 1.5cm 0.1cm},clip,width=\columnwidth,height=5.5cm]{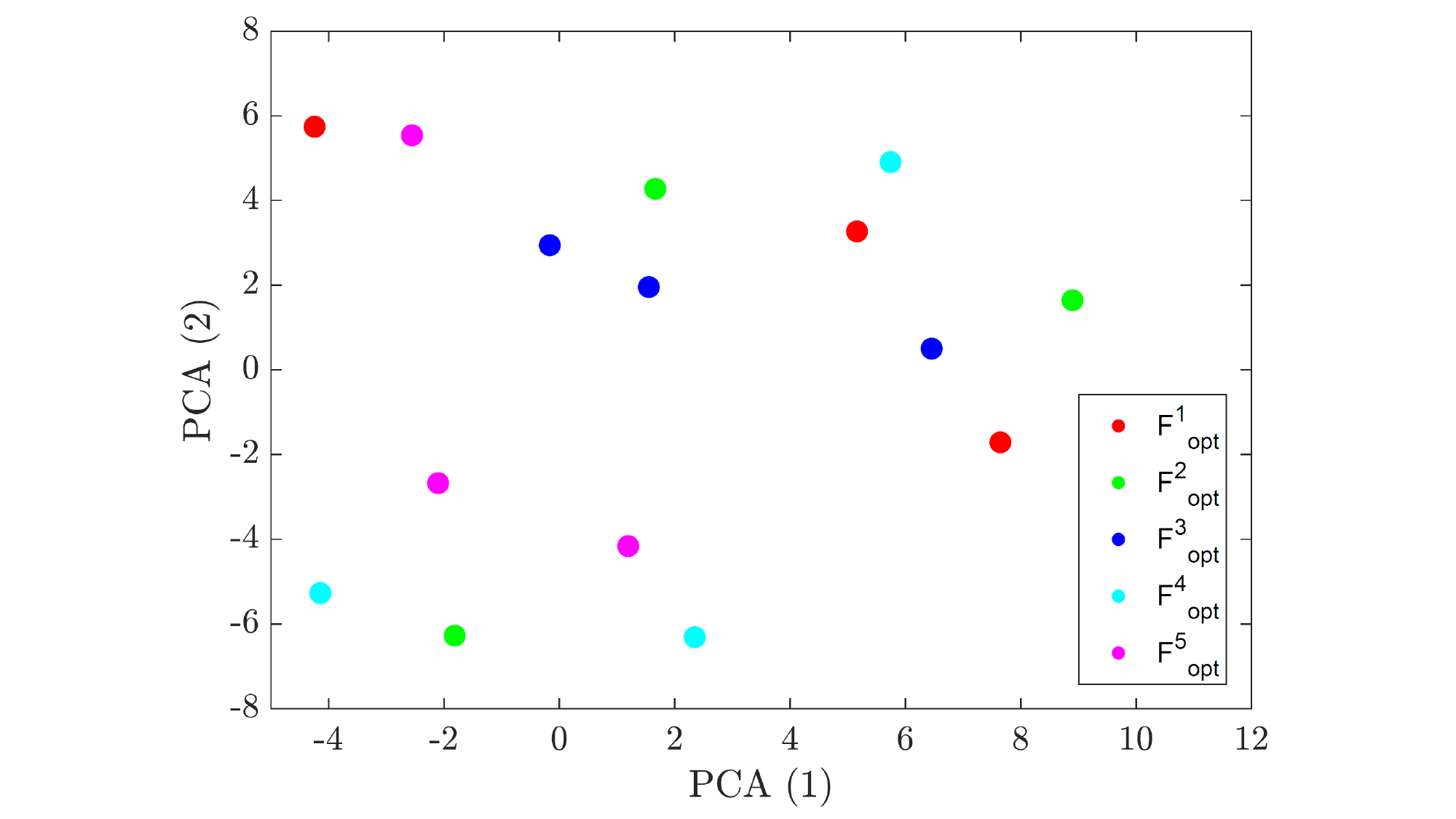}
  %\vspace{-15pt}
    \caption{}
    \label{fig:embeddings_before}
     \end{subfigure}
     \hfill
     \begin{subfigure}[b]{0.5\textwidth}
         \centering
         \includegraphics[height=5.5cm,trim={1.5cm 0.1cm 1.6cm 0.1cm},clip,width=\columnwidth]{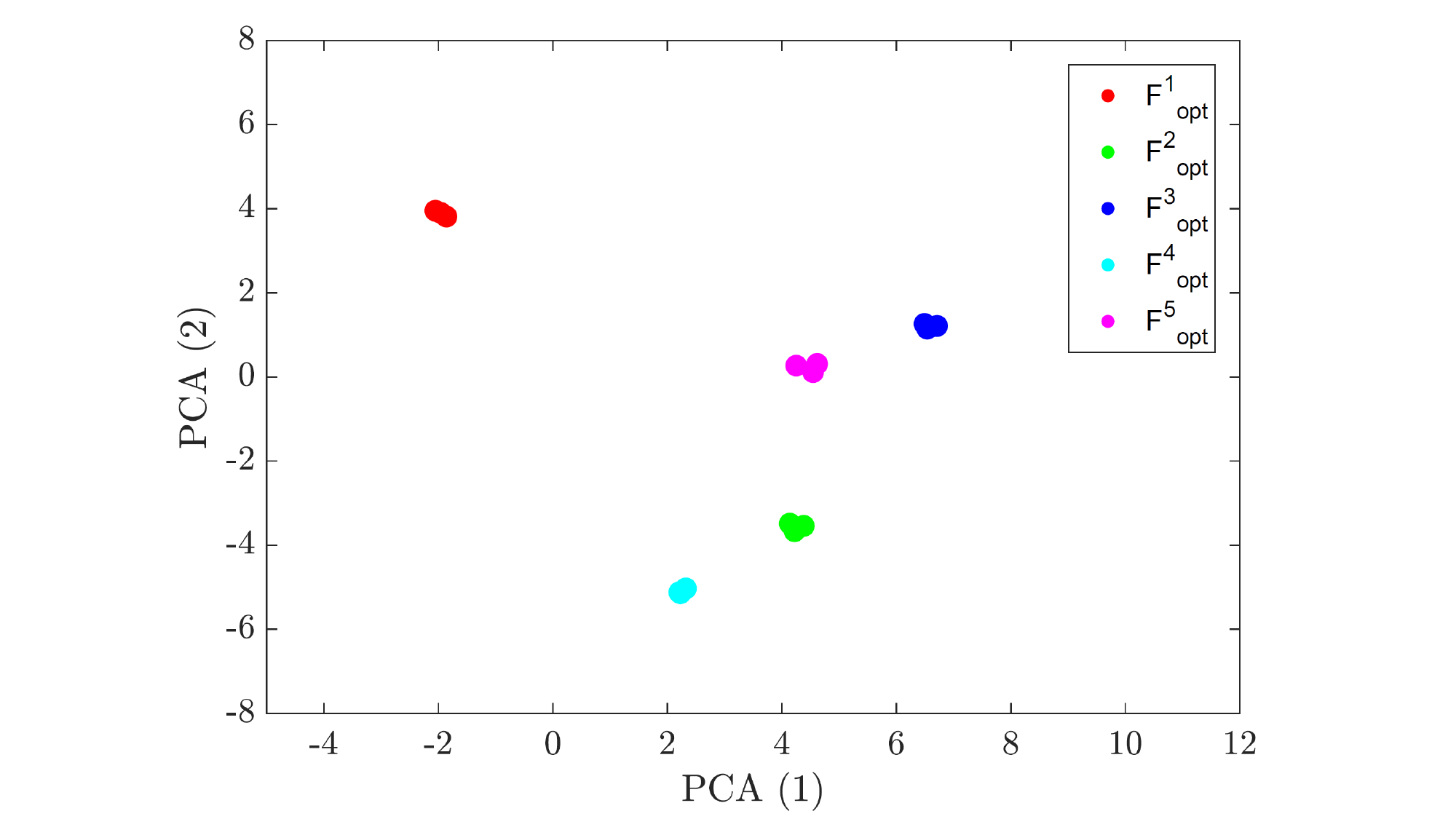}
    \caption{}
    \label{fig:embeddings_after} 
     \end{subfigure}
\caption{PCA plot of $\RM{F}_{\mathrm{opt}}$ embeddings illustrating the model's ability to cluster augmented samples from the same channel state: (a) before and (b) after contrastive pre-training.}
  \label{fig:embeddings}
  \vspace{-4mm}
\end{figure}

\subsection{Contrastive Learning for Robust UM-MIMO}

Leveraging self-supervised contrastive learning, our approach aims to enhance hybrid beamforming performance under imperfect CSI. We begin with a batch of optimal beamforming matrices $\RM{F}_{\mathrm{opt}}$ as a base for augmentation. By adding AWGN noise\footnote{Our methodology is universally applicable to various noise models.}, we generate two augmented samples, $\Tilde{\RM{F}}_{\mathrm{opt}}$, for each $\RM{F}_{\mathrm{opt}}$, creating positive pairs, while the rest unrelated matrices within the same batch are treated as negative samples. This generates a rich dataset that captures the variance in channel conditions due to noise. Our contrastive loss function then works to cluster the $D_E$ dimensional embeddings (as depicted in Fig. \ref{fig:cont_block}) of augmented samples from the same channel state, driving the model to learn robust representations despite noisy inputs.

 Fig. \ref{fig:embeddings} presents principal component analysis (PCA) plots of the embeddings before and after contrastive training. We select five samples for visualization and generate two additional augmented samples for each by adding noise. Initially, Fig. \ref{fig:embeddings}(a) reveals significant dispersion among the original and their corresponding augmented samples, reflecting the impact of the noise. Post-contrastive training (Fig. \ref{fig:embeddings}(b)), the PCA plot indicates that the original and augmented samples' embeddings have converged, demonstrating our training method's efficacy in creating tightly clustered representations resilient to CSI noise. These refined embeddings directly enable our model to learn how to neutralize CSI-induced noise impacts adeptly. This equivalence in embeddings allows for generating accurate beamforming matrices under noisy conditions, mirroring the precision achievable with perfect CSI.

\begin{figure*}[!ht]
\centering
    \includegraphics[width=2\columnwidth,height=80mm,trim={0.6cm 2.8cm 0.1cm 1.1cm},clip]{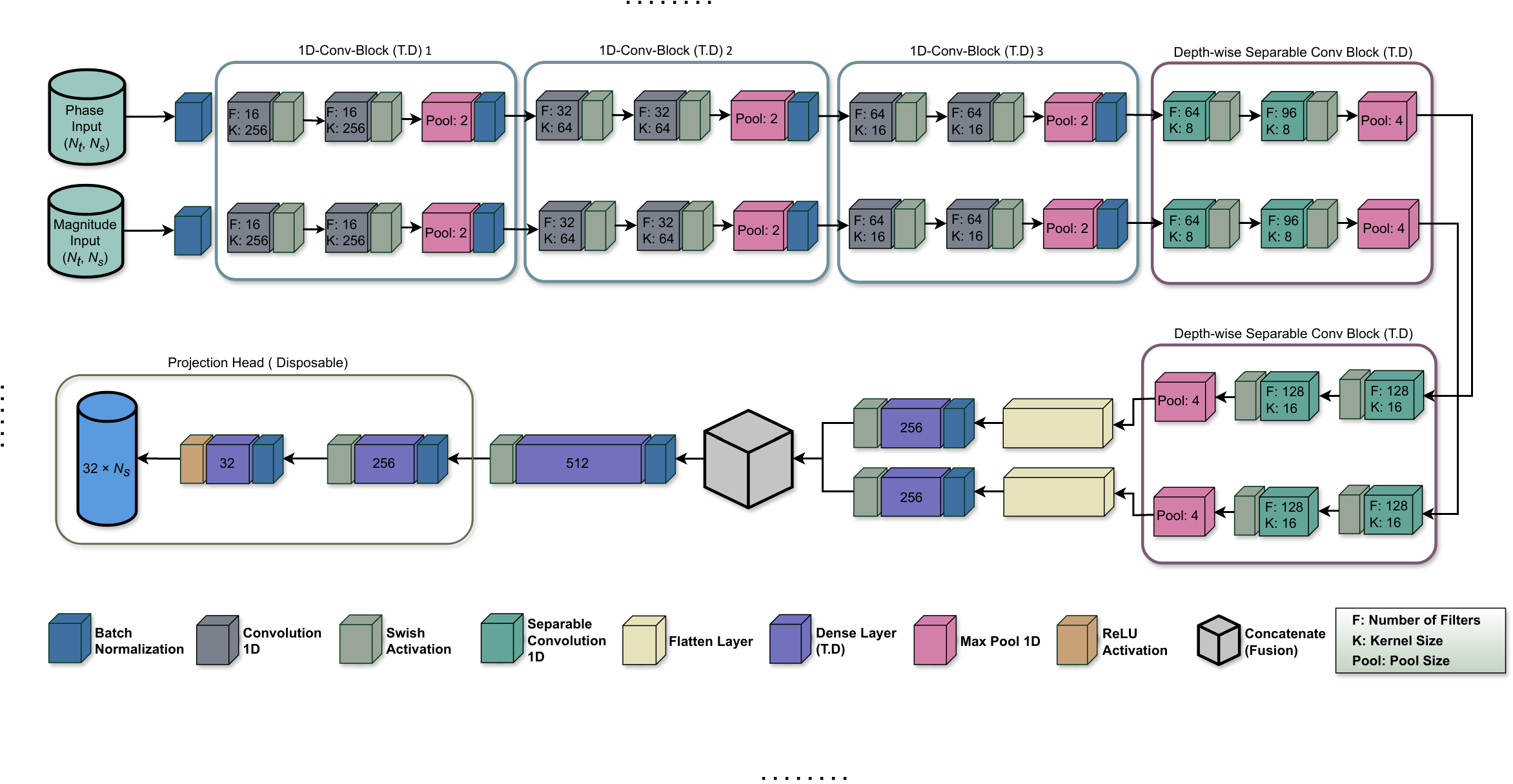}
    \caption{Detailed view of the contrastive pre-training network showcasing the projection head for embedding generation.}
    \vspace{-2mm}
    \label{fig:cont_archi}
\end{figure*}

\begin{figure*}[!ht]
\centering
    \includegraphics[width=1.9\columnwidth,height=60mm,trim={8cm 16cm 5cm 4.5cm},clip]{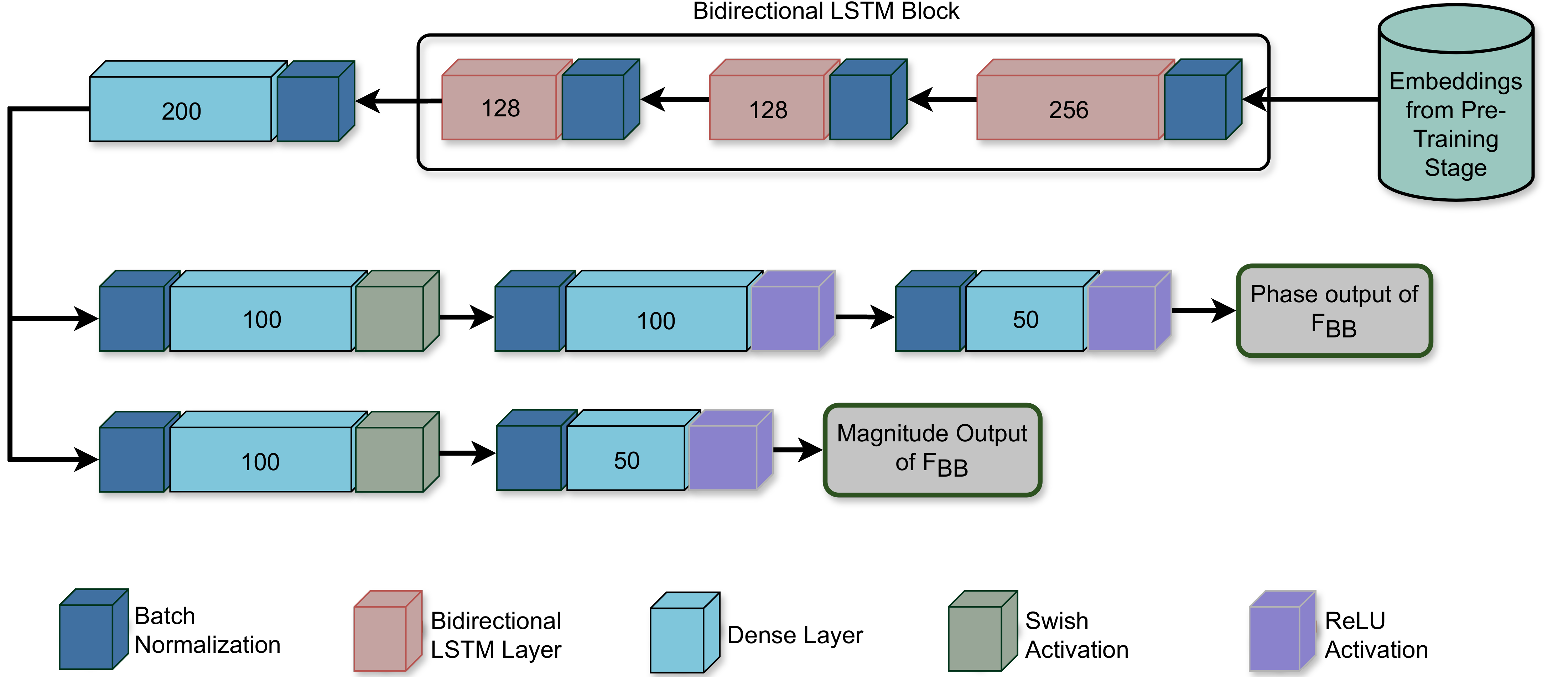}
    \caption{Architecture of the final-stage training network with bidirectional LSTM blocks for generating the beamforming matrix.}
    \vspace{-2mm}
    \label{fig:final_archi}
\end{figure*}

\subsection{Architecture of the Proposed Contrastive Network}

Our contrastive pre-training network employs a multi-branch architecture to independently process phase and magnitude inputs of beamforming matrices before fusion as depicted in Fig. \ref{fig:cont_block}. Each branch consists of a series of 1D convolutional (Conv) blocks, with time-distributed (T.D) operations to handle the temporal aspects of our data. The network accepts $\RM{F}_{\mathrm{opt}}$ phase and magnitude inputs of dimension ($N_{t}$,$N_{s}$).

The first Conv block in each branch includes separable convolutional layers with 16 filters of kernel size 256, succeeded by batch normalization and ReLU activation. Subsequent blocks increment the filter count and decrement the kernel size, intensifying feature specificity. After three Conv blocks, depth-wise separable convolutional layers aim to learn spatial hierarchies efficiently. Max-pooling layers after Conv blocks reduce dimensionality and highlight key features.

Following feature extraction, branches merge, forming a fusion network that advances through a projection head. This projection head, discardable after training, comprises a shallow multilayer perceptron (MLP) with a flatten layer, ensuring a succinct $D_P$ dimensional representation of the $D_E$ embeddings for the contrastive loss function. The architecture is meticulously designed to ensure the embeddings from the phase and magnitude of $\RM{F}_{\mathrm{opt}}$ converge, facilitating effective contrastive learning across diverse channel conditions. The detailed architecture of the contrastive pre-training network is shown in Fig. \ref{fig:cont_archi}.

\subsection{Beamforming Matrix Generation}

Our network's final stage of training employs a novel architecture specifically designed for generating beamforming matrices. The network architecture comprises a bidirectional Long Short-Term Memory (LSTM) block with multiple LSTM layers that process the embeddings generated from the contrastive stage, as shown in the prediction head of Fig. \ref{fig:cont_block}.

The bidirectional LSTM block, with 256, 128, and 128 units in each subsequent layer, respectively, ensures the capture of temporal dependencies within the data. This is essential for learning the dynamics within the phase and magnitude representations of $\RM{F}_{\mathrm{BB}}$. The output from the LSTM is concatenated, resulting in a feature vector of size 200, which encapsulates temporal features from both directions.

Concurrently, the network's lower branch processes the magnitude and phase separately. Each path features a stack of dense layers with 100 neurons, followed by a reduction to 50. These dense layers, selected for efficiency, are accompanied by swish activations to emphasize salient features. The outputs are then flattened and input into subsequent dense layers. The detailed architecture of the contrastive pre-training network is displayed in Fig. \ref{fig:final_archi}.

The LSTM and lower branch outputs work by combining temporal and spatial features. This composite feature set is then refined through additional dense layers, culminating in the final magnitude and phase outputs of $\RM{F}_{\mathrm{BB}}$, each with dimensions \( (N_\mathrm{RF}, N_s) \). This architecture facilitates a sophisticated learning mechanism for predicting beamforming matrices adept at navigating the complexities of real-world UM-MIMO systems.

In this architecture, the self-supervised characteristic emerges from the fact that the input $\RM{F}_{\mathrm{opt}}$ is also used to calculate the loss function, which is formulated analogously to (2). Here $\RM{F}_{\mathrm{RF}}$ is obtained via the product of the input $\RM{F}_{\mathrm{opt}}$ and pseudo-inverse of the generated output $\RM{F}_{\mathrm{BB}}$. This innovative approach underpins the training process, ensuring that the network's output, $\RM{F}_{\mathrm{BB}}$, is consistently aligned with the optimal beamforming matrices. By integrating the input as part of the loss computation, the network is intrinsically guided toward solutions that inherently respect the underlying physical model of UM-MIMO systems.

\begin{figure}[t]
    \includegraphics[width=\linewidth,height=65mm,trim={3.5cm 0.6cm 3.5cm 0.7cm},clip]{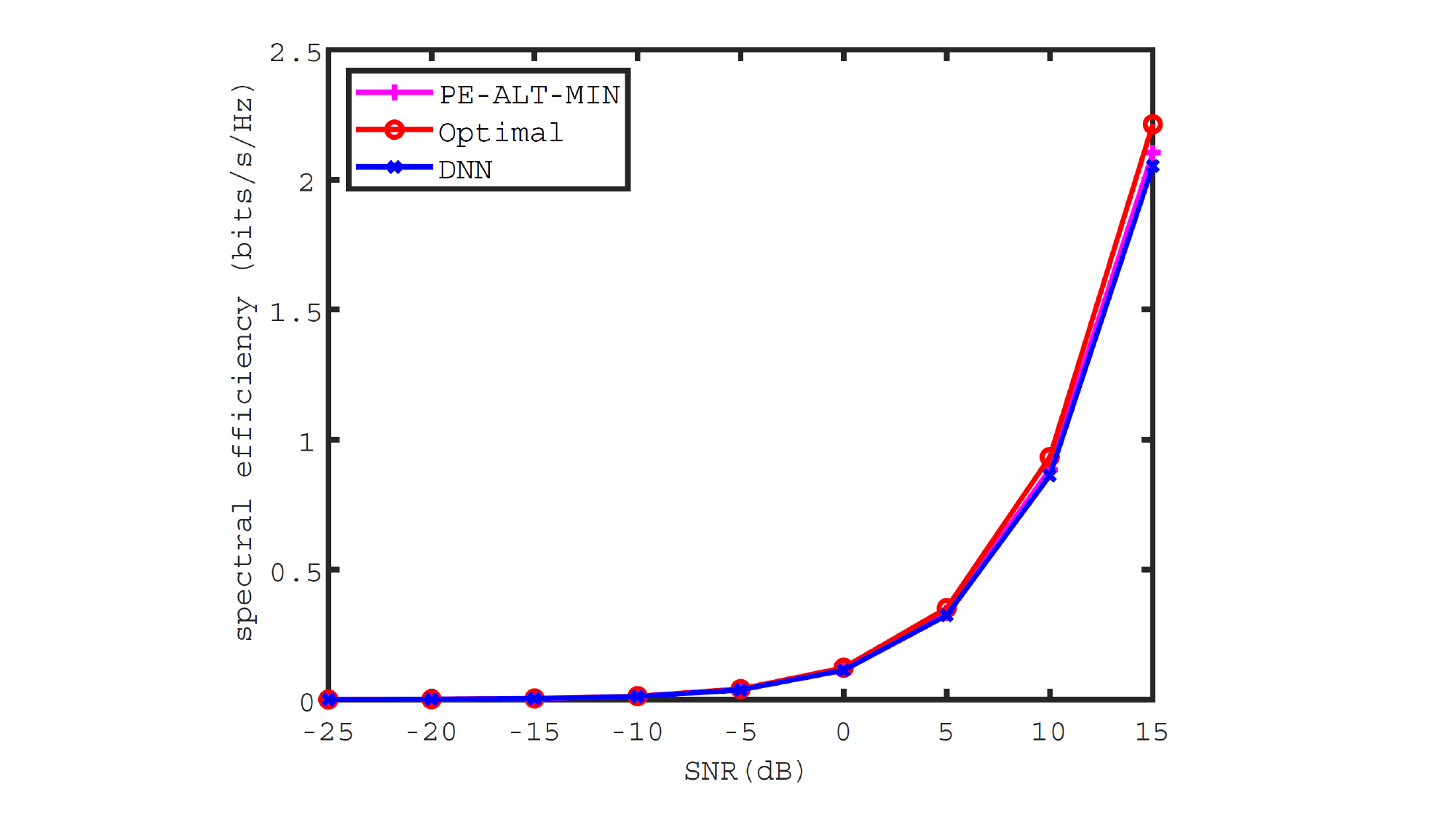}
    \caption{SE as a function of SNR for perfect CSI scenarios.}
    \vspace{-4mm}
    \label{fig:perfect_CSI}
\end{figure}

\section{Simulation Results and Performance Analysis}
We evaluate the performance of our proposed approach through extensive simulations under various imperfect CSI conditions. We compare our proposed method's achievable rate with existing hybrid beamforming techniques and show that our contrastive learning-based approach outperforms traditional methods, particularly under severe CSI imperfections.

\subsection{Environment}

Our framework has been developed utilizing Python, Tensorflow 2.12, and Keras 2.12. Numpy is employed for computations, while MATLAB 2023b is utilized for generating certain visual representations. The model is trained on Google Colab Pro+ using a system with 83.5 GB of RAM, 166.8 GB of disk space, and an NVIDIA A100-SXM4-40GB GPU to accelerate the training process. After training, the model's size is compact ($\thickapprox$ 1.3 MB), enabling it to be deployed on regular PC hardware and edge computing platforms.

\begin{table}[h]
    \vspace{3mm} \caption{NYUSIM Simulation Parameters \vspace{-1mm}}
    \label{nyusim_channel}
    \renewcommand{\arraystretch}{1.2}
    \providecommand{\cellLeft}[1]{\hspace{0.5cm}#1}
    \providecommand{\cellRight}[1]{\hspace{1.0cm}#1}
    \centering
    \begin{tabular}{l p{4.0cm}}
        \hline
        \cellLeft{\textbf{Parameter}} & \cellRight{\textbf{Value}} \\
        \hline
        \rowcolor{blue!10!white} 
        \cellLeft{Scenario} & \cellRight{UMa} \\
        \cellLeft{Frequency} & \cellRight{100 GHz} \\
        \rowcolor{blue!10!white} 
        \cellLeft{RF Bandwidth} & \cellRight{800 MHz} \\
        \cellLeft{TX and RX Antenna Type} & \cellRight{URA} \\
        \rowcolor{blue!10!white} 
        \cellLeft{Number of TX Antennas} & \cellRight{1024} \\
        \cellLeft{Number of RX Antennas} & \cellRight{16} \\
        \rowcolor{blue!10!white} 
        \cellLeft{Environment} & \cellRight{LOS} \\
        \cellLeft{T-R Separation Distance} & \cellRight{$10 - 30$ m} \\
        \rowcolor{blue!10!white} 
        \cellLeft{TX Power} & \cellRight{50 dBm} \\
        \cellLeft{Base Station Height} & \cellRight{35 m} \\
        \rowcolor{blue!10!white} 
        \cellLeft{User Terminal Height} & \cellRight{1.5 m} \\
        \cellLeft{Barometric Pressure} & \cellRight{1013.25 mbar} \\
        \rowcolor{blue!10!white} 
        \cellLeft{Humidity} & \cellRight{50} \% \\
        \cellLeft{Temperature} & \cellRight{20 $^{\circ}$C} \\
        \rowcolor{blue!10!white} 
        \cellLeft{Rain Rate} & \cellRight{0.1 mm/hr} \\
        \hline
    \end{tabular}
    \vspace{-3mm}
\end{table}

\begin{figure}[t]
    \includegraphics[width=\linewidth,height=65mm,trim={4cm 0.2cm 3.5cm 0.8cm},clip]{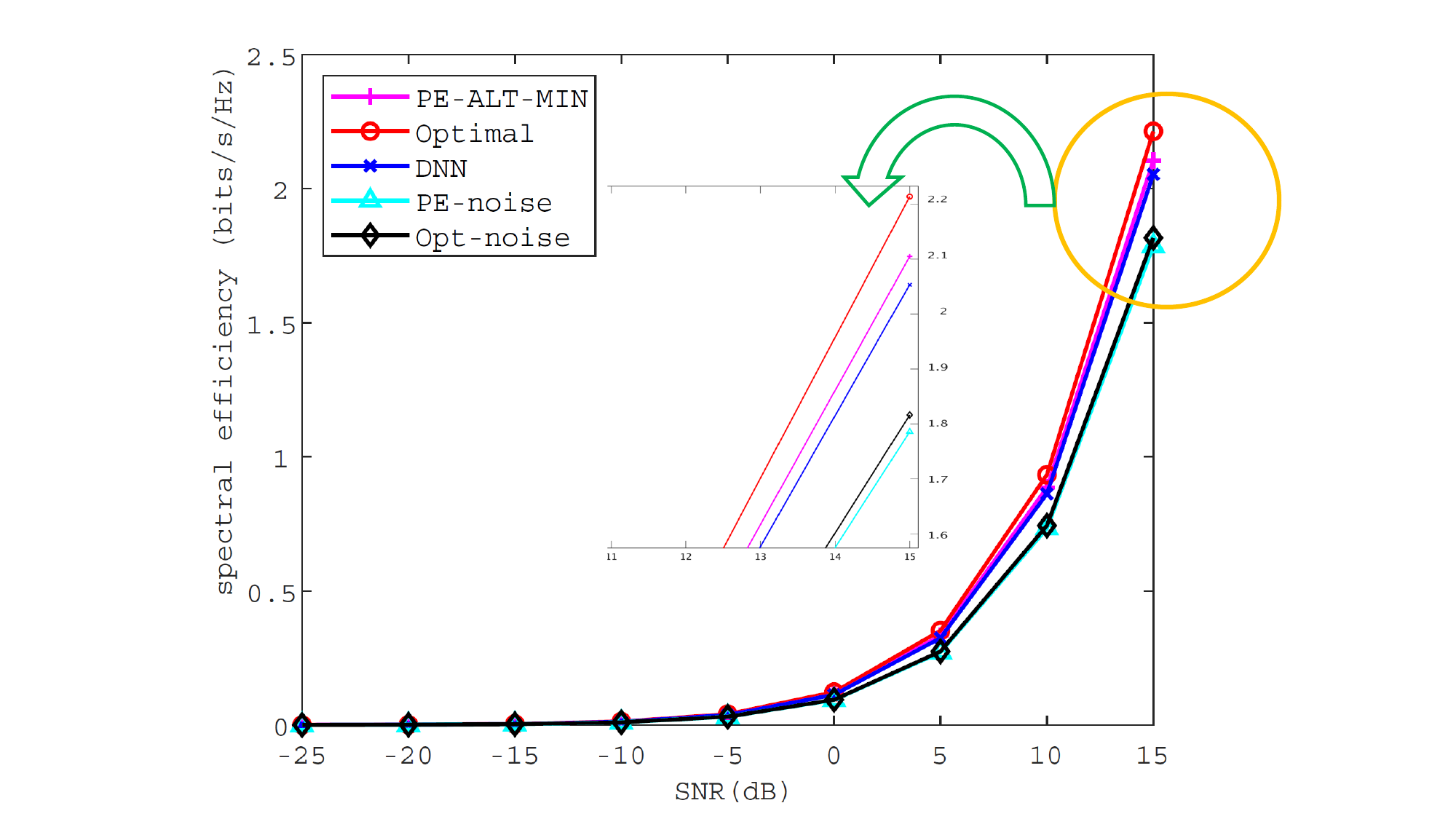}
    \caption{Comparison between achieved SE under perfect CSI and imperfect CSI at 36 dB CSI SNR.}
    \vspace{-4mm}
    \label{fig:CSI_36}
\end{figure}

\begin{figure*}[ht]
\centering
    \begin{subfigure}[b]{0.49\linewidth}
        \includegraphics[width=\linewidth,height=65mm,trim={2.7cm 0.1cm 2.0cm 0.7cm},clip]{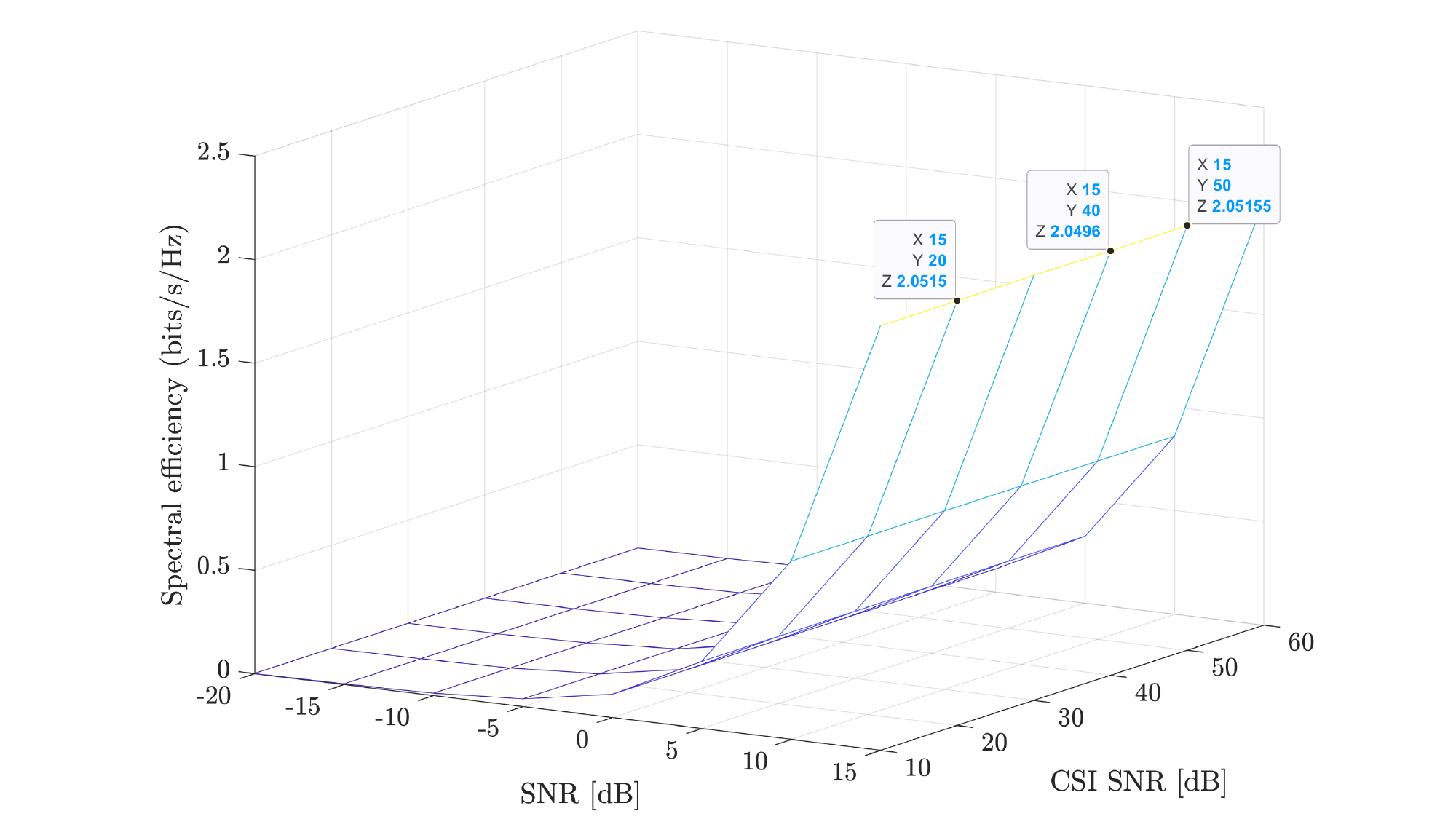}
        \caption{Proposed DNN}
        \label{fig:imperfectCSI_DNN}
    \end{subfigure}
    \hfill
    \begin{subfigure}[b]{0.49\linewidth}
        \includegraphics[width=\linewidth,height=75mm,trim={2.2cm 0.1cm 2.0cm 0.5cm},clip]{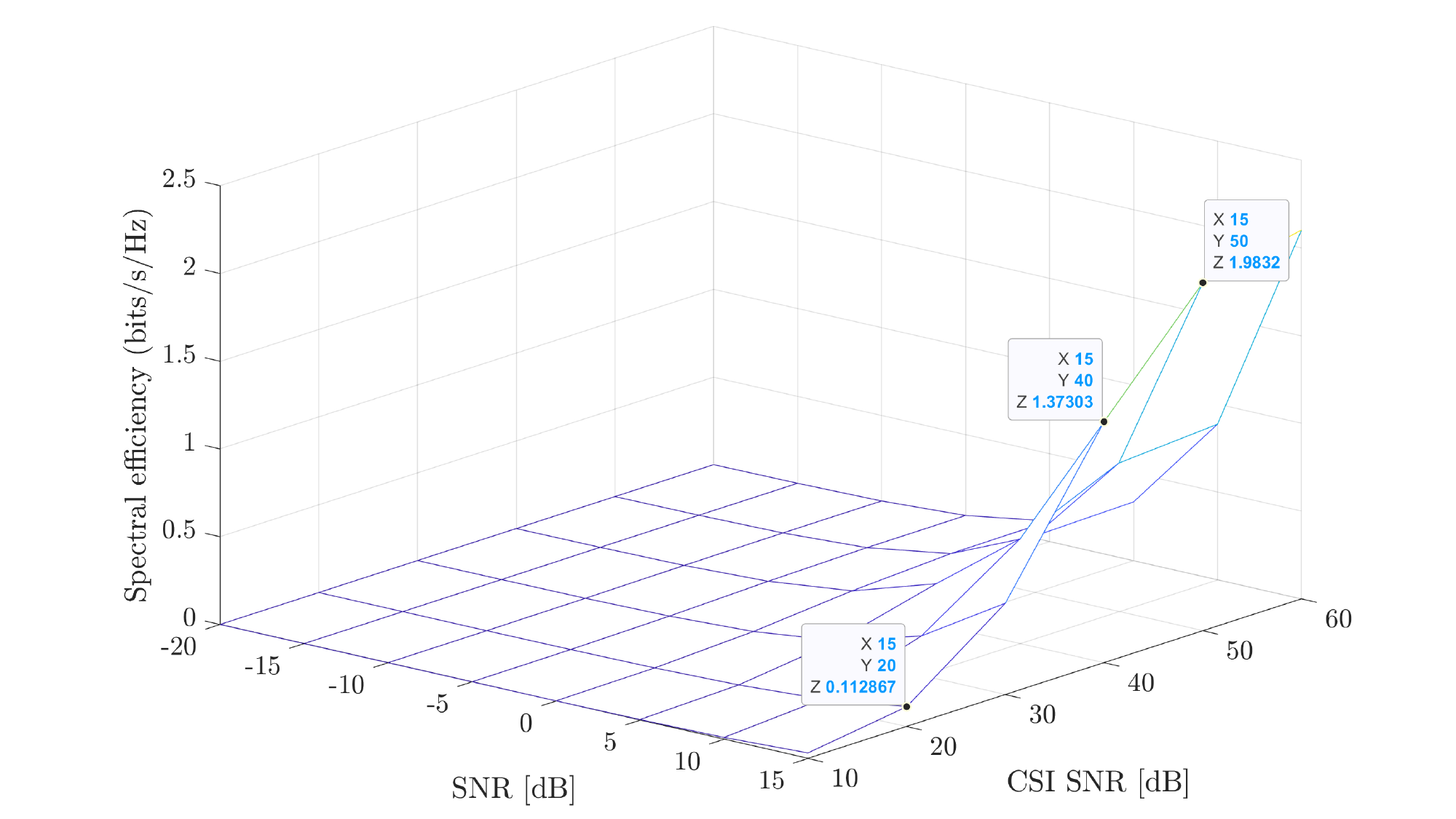}
        \caption{PE-ALT-MIN}
        \label{fig:imperfectCSI_PE_ALT_MIN}
    \end{subfigure}
    \caption{SE as a function of SNR under different CSI imperfections.}
    \vspace{-3mm}
    \label{fig:imperfectCSI}
\end{figure*}

\subsection{Data Generation}
The NYUSIM channel simulator (version 3.1) is utilized for generating channel data with a carrier frequency of 100 GHz and an RF bandwidth of 800 MHz\cite{zekri2019study}. We contemplate a uniform rectangular array, square in shape, comprising 1024 elements for the transmitter and a receiver array with 16 elements, both featuring a standard element spacing of 0.5 wavelength. We consider a rain rate of 0.1 mm/hr to incorporate rain fade and emulate real-life scenarios\cite{budalal2021millimetre}. The parameters for dataset generation are outlined in Table \ref{nyusim_channel}. 

To train and validate our proposed deep neural network (DNN) model, we generate 50,000 channel realizations using the previously specified configuration. We also produce an additional 10,000 channel realizations (test set) to evaluate the efficacy of the proposed model.

\subsection{Performance under Perfect CSI}

Before discussing our model's efficacy under imperfect CSI, it is illustrative to consider its performance with perfect CSI knowledge, serving as a benchmark for the universality of our approach. The results exhibit a promising alignment with the theoretical optimum in assessing our model's performance under perfect CSI conditions. The graph shown in Fig. \ref{fig:perfect_CSI} plots SE against signal-to-noise ratio (SNR), where our proposed DNN approach and the baseline Alternating Minimization (PE-Alt-MIN) method \cite{yu2016alternating} are compared against the optimal beamforming performance. Here, the optimal digital beamformer is obtained directly via the SVD of the channel matrix. As SNR increases, both the proposed DNN and PE-Alt-MIN closely track the optimal performance, merging with the optimal curve at most SNR values. This indicates that our approach achieves an identical SE to Alt-MIN under perfect CSI, as shown by the overlapping lines on the graph. The simulation reinforces the DNN's capacity to approximate optimal beamforming techniques, validating the efficacy of our model in ideal conditions where we presume an optimal analog beamformer is in place.

\subsection{Performance under Imperfect CSI}

To examine our model's resilience under imperfect CSI, we first compare its performance at a CSI SNR of 36 dB (close to perfect CSI) and under perfect CSI conditions. The results are presented in Fig. \ref{fig:CSI_36}, where the plots also show the performance of the benchmark PE-Alt-MIN and the optimal digital beamformer under both perfect CSI and noisy CSI (36dB). The SE of our DNN approach maintains a near-constant level, demonstrating robustness even as the CSI SNR varies. This is contrasted with the Alt-MIN algorithm, which exhibits a significant drop in performance with decreasing CSI SNR. Here, the SE achieved by the DNN in both perfect CSI conditions and 36dB CSI SNR remains exactly the same, and hence, only the latter is shown in the plot. The numerical results reveal a gain of more than 10\% in SE compared to other methods, affirming the DNN's effectiveness. 

For a more comprehensive performance analysis, we vary the extent of CSI imperfection and examine the SE achieved by the algorithms in such scenarios. As further illustrated in Fig. \ref{fig:imperfectCSI}(a), the SE of our DNN remains stable and performs comparably to the optimal beamforming scenario, signifying the model's insensitivity to CSI imperfections. In comparison, the SE of Alt-MIN, as depicted in Fig. \ref{fig:imperfectCSI}(b), reduces drastically with the increment of noise in CSI, emphasizing the DNN's superior noise resilience. Particularly, the SE of Alt-MIN decreases by almost 17 times at a CSI SNR of 20 dB, indicating a steep performance degradation under increased noise levels. This drastic performance decline in Alt-MIN can be attributed to the unique challenges UM-MIMO systems pose. The relatively fewer multipath components and the highly focused beams formed by the extensive antenna arrays demand unparalleled precision in beamforming strategies. Such precision is elusive for algorithms like Alt-MIN under noisy CSI conditions. Our DNN model's ability to maintain high SE across diverse CSI SNR levels demonstrates its adaptability and suitability for real-world UM-MIMO applications, where it successfully navigates the nuances of channel imperfections.

\section{Conclusion}
In this paper, we have proposed a novel contrastive learning-based approach for hybrid beamforming in UM-MIMO systems under imperfect CSI conditions. The efficacy of the proposed method is substantiated through numerical results, which reveal a marked improvement in system performance—specifically in achievable rate—over conventional techniques. Notably, at a CSI SNR of 20 dB, our approach maintains stable SE, in stark contrast to traditional beamforming algorithms, which exhibit a seventeenfold decrease in SE. Additionally, our unique approach showcases potential in handling non-instantaneous CSI challenges, enhancing adaptability for user mobility and dynamic environments, thereby paving the way for addressing key 6G UM-MIMO communication issues, including energy-efficient resource allocation.

\bibliographystyle{IEEEtran}
\bibliography{bibtex/bib/IEEEabrv,bibtex/bib/main_letter}

\end{document}